\documentclass[3p,times,procedia]{elsarticle}
\usepackage{nupha_ecrc}

\volume{00}

\firstpage{1}

\journalname{Nuclear Physics A}

\runauth{}

\usepackage{amssymb}

\usepackage[figuresright]{rotating}

\begin{document}

\begin{frontmatter}



\dochead{}

\title{Heavy flavour production in proton-lead and
lead-lead collisions with LHCb}


\author{Michael Winn}

\address{Laboratoire de l'Acc\'el\'erateur Lin\'eaire}


\begin{abstract}
  The LHCb experiment offers the unique opportunity to study heavy-ion interactions in the forward region (2  $<\eta<$  5), in a kinematic
domain complementary to the other 3 large experiments at the LHC. The detector has excellent capabilities for reconstructing quarkonia and open charm states, including baryons, down to zero $p_T$. It can separate the prompt and displaced charm components. In $p$Pb collisions, both forward and backward rapidities are covered thanks to the possibility of beam reversal. Results include measurements of the nuclear modification factor and forward-backward ratio for charmonium, open charm and bottomonium states. These quantities are sensitive probes for nuclear effects in heavy flavour production. Perspectives are given with the large accumulated luminosity during the 2016 $p$Pb run at the LHC.   In 2015, LHCb participated successfully for the first time in the PbPb data-taking. The status of the forward prompt J/$\psi$ nuclear modification factor measurement in lead-lead collisions is discussed.
\end{abstract}


\end{frontmatter}


\vspace{0.4cm}

Measurements of heavy-flavour 
production are important tools to investigate nuclear effects both in proton-lead and in lead-lead collisions~\cite{Andronic:2015wma}. The large mass sets an infrared cut-off and therefore allows to calculate cross sections down to zero transverse momentum with perturbative QCD (pQCD). Furthermore, they are produced at an early stage of the collision.
In nucleus-nucleus collisions,  the produced charm or beauty quarks are hence sensitive to the full evolution of the thermodynamic system. Their kinematic distributions and their fragmentation into hadrons can change compared to the behaviour in $pp$ collisions. The properties of the bound states of charm and beauty quark pairs are sensitive to deconfinement and their production encodes therefore valuable information about the created system in the laboratory~\cite{Mocsy:2013syh}. In proton-nucleus collisions, the investigation of applicability and the improvement of nuclear parton distribution functions, e.g.~\cite{Eskola:2009uj,Kovarik:2015cma,Eskola:2016oht}, Color Glass Condensate approach calculations~\cite{Fujii:2006ab}, coherent energy loss mechanism~\cite{Arleo:2012rs} and other effects not related to deconfinement are at the centre of the measurement interest in order to provide a better understanding of pQCD and in order to provide input for the understanding of nucleus-nucleus collisions.

LHCb is well suited to measure charm and beauty production not only in $pp$ collisions~\cite{LHCb-DP-2014-002}, but also in $p$Pb and in peripheral PbPb collisions. The detector is a fully instrumented single arm forward spectrometer (2$<\eta<$5) equipped with a 4 Tm dipole magnet. In particular, it features a silicon strip detector, the VErtex LOcator (VELO), that covers the radial distance from  the beam between 8 and 42~mm. 
Tracker stations upstream and downstream of the  magnet enable a precise measurement of the momentum together with the VELO detector.  Two RICH detectors allow charged particle identification in the momentum range between 10 to 100~GeV/$c$. An electromagnetic calorimeter, a hadronic calorimeter and a muon system for particle identification, photon detection and hardware triggers complete the detector set-up.

\begin{figure}
  \includegraphics[width=0.45\textwidth]{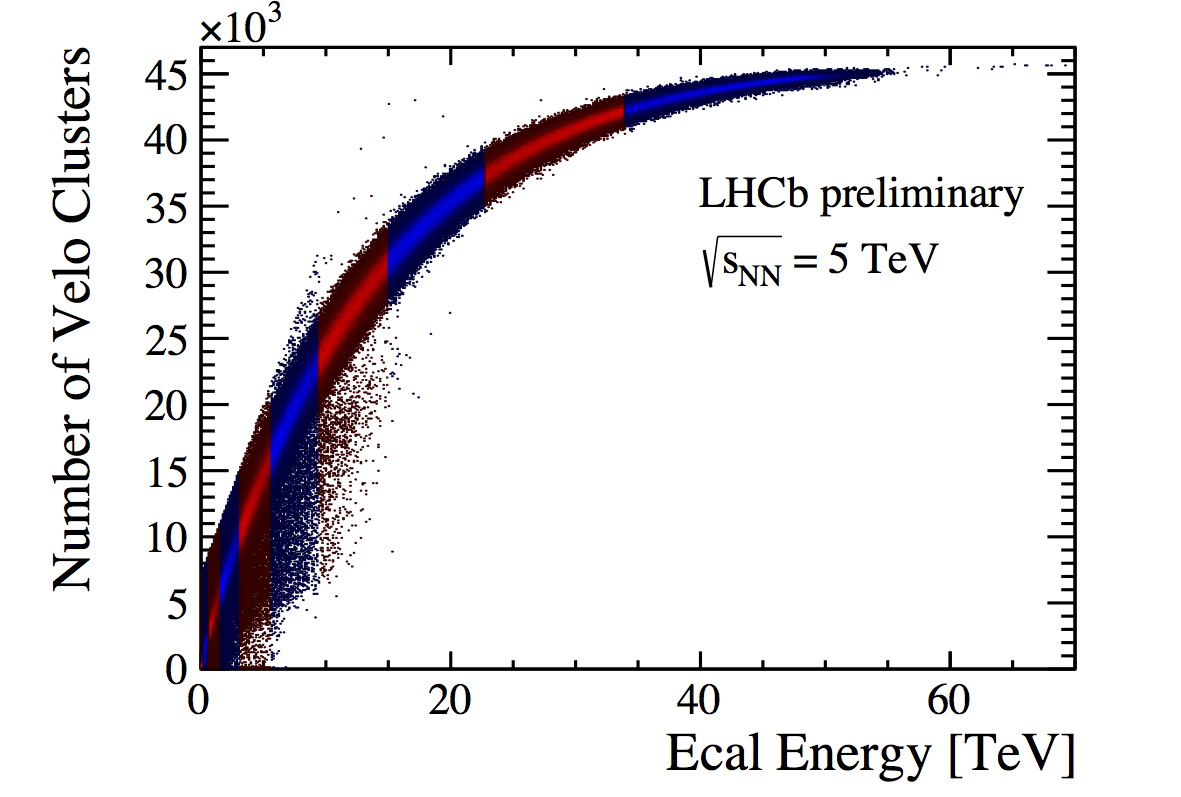}
  \includegraphics[width=0.45\textwidth]{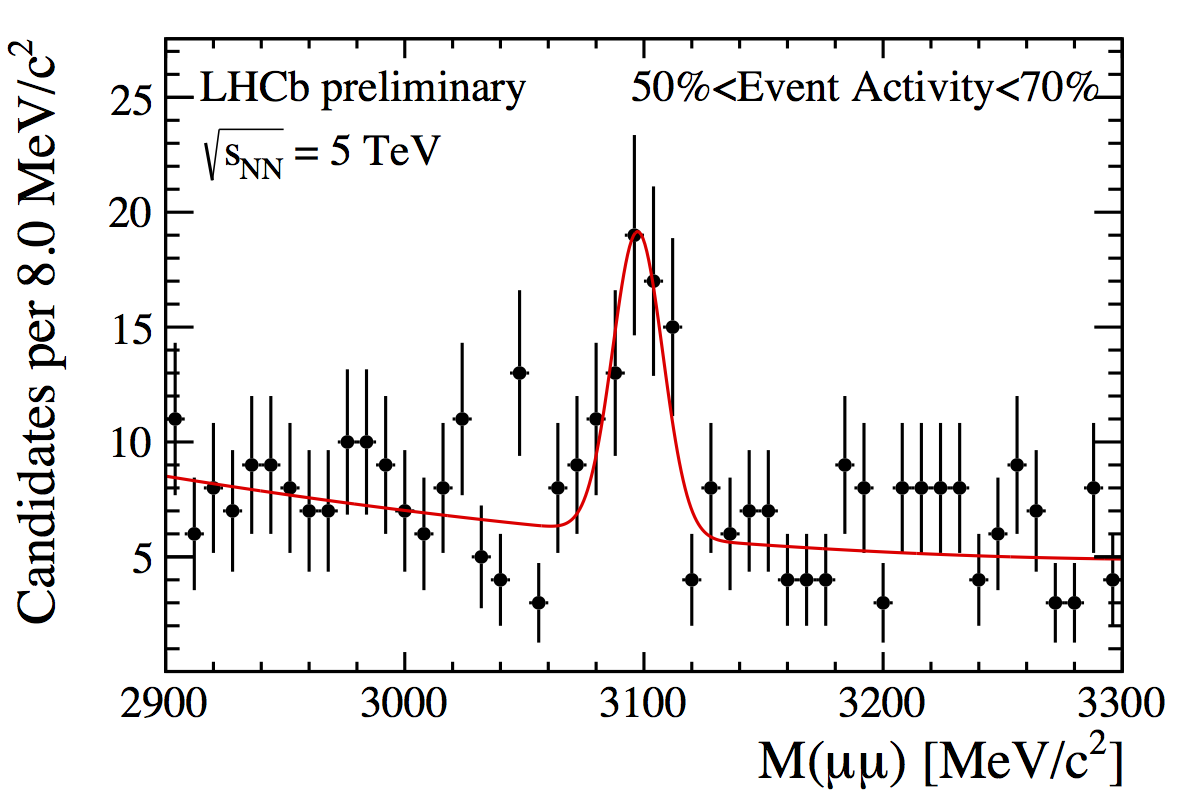}
  \caption{LHS: relation between clusters in the VELO detector, part of the tracking system, and the energy deposited in the electromagnetic calorimeter (Ecal) in PbPb collisions. The color bands indicate 10\% centrality intervals measured with the Ecal. The bending of the band signals the saturation of the tracking detector. RHS: The dimuon candidate invariant mass distribution in the vicinity of the J/$\psi$ mass in the centrality range 50-70\%. A clear signal is visible.  }
  \label{fig:PbPb}
  \end{figure}

In 2015, LHCb collected for the first time PbPb collisions. The detector  recorded about 50 million minimum bias collisions with a data taking efficiency of around 90\% excluding dead-time. The detector is designed for  running in $pp$ collisions at an average number of visible interactions of order one. The limitations of the detector in central PbPb collisions  can be seen in Fig.\ref{fig:PbPb} on the left hand side:  it becomes evident that the tracker occupancy saturates at 100\% for most central collisions. With the current tracking algorithms, it is possible to reconstruct events up to a centrality of about 50\%. An analysis is ongoing aiming at the extraction of the nuclear modification factor of prompt J/$\psi$ production in peripheral collisions.   The signal extraction of the J/$\psi$ is shown for the centrality interval 50-70\% in Fig.~\ref{fig:PbPb} on the right hand side. A mass resolution of about 10 MeV/$c^2$ is achieved similar to the resolution in $pp$ collisions. The analysis efforts concentrate in PbPb at the moment on the tracking efficiency evaluation with data-driven methods. 

\begin{figure}
  \includegraphics[width=0.45\textwidth]{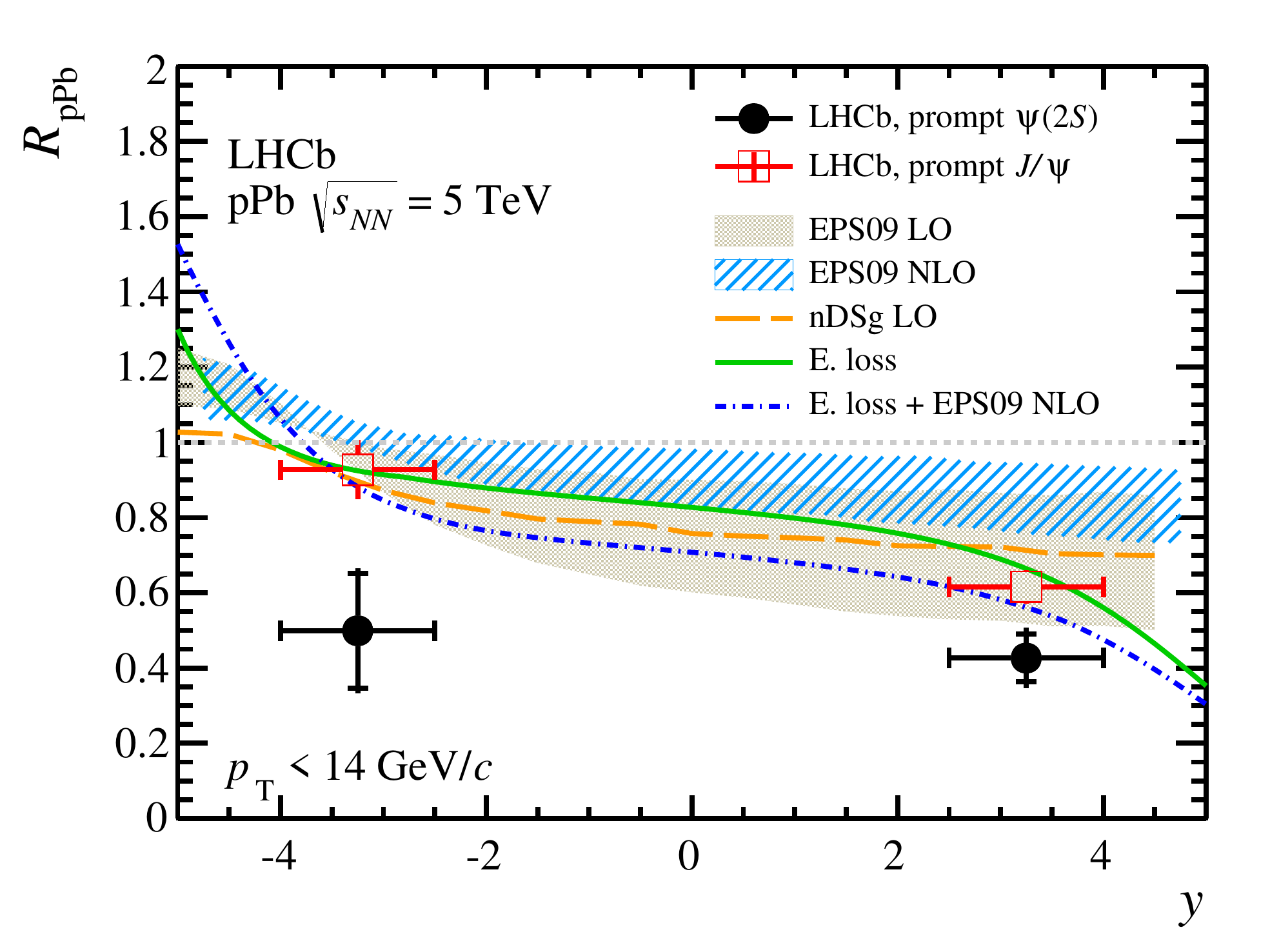}
  \includegraphics[width=0.45\textwidth]{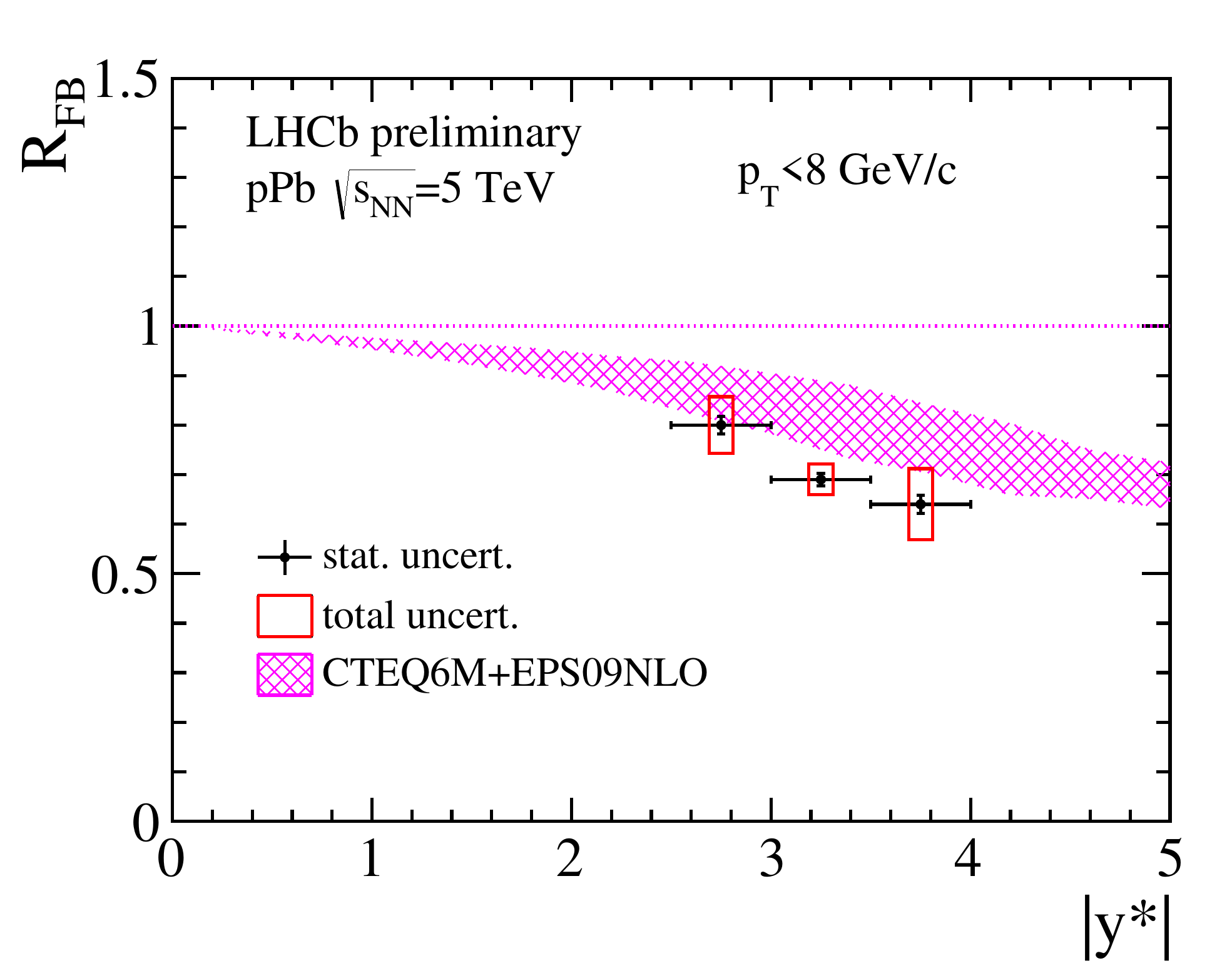}
  \caption{LHS: The nuclear modification factor of J/$\psi$ and $\psi$(2S) compared with models: Ref.~\cite{Ferreiro:2013pua} (yellow dashed line and brown band), Ref.~\cite{Albacete:2013ei} (blue band),  Ref.~\cite{Arleo:2012rs} (green solid and blue dash-dotted lines). A significant additional suppression of the $\psi$(2S) is visible. RHS: the forward to backward ratio of $D^0$ production in $p$Pb collisions. The measurement is compared with a calculation based on NLO pQCD hard-matrix elements~\cite{Mangano:1991jk}, the nuclear PDF set EPS09~\cite{Eskola:2009uj} and the pp PDF set CTEQ6M~\cite{Stump:2003yu}.}
  \label{fig:pPb13}
  \end{figure}

LHCb participated successfully in the $p$Pb data taking 2013 at $\sqrt{s_{NN}}=$5~TeV. The forward acceptance implies coverage down to Bjorken-$x$ values of $10^{-6}$ with heavy quarks, a regime, where no experimental data exists prior to the LHC in nuclear collisions. In Fig.~\ref{fig:pPb13}, the nuclear modification factor of prompt J/$\psi$~\cite{LHCb-PAPER-2013-052} and prompt $\psi$(2S)~\cite{LHCb-PAPER-2015-058} is shown compared with different phenomenological calculations.  Collinear pQCD calculations using nuclear parton distribution functions~\cite{Albacete:2013ei,Ferreiro:2013pua} can describe the data within uncertainties as well as the coherent energy loss model~\cite{Arleo:2012rs}. Color Glass Condensate calculations published after the publication of the experimental data in the dilute-dense limit are also able to describe the data at forward rapidity~\cite{Ma:2015sia,Ducloue:2015gfa,Ducloue:2016pqr}. However, the behaviour of the excited $\psi$(2S) bound state production, an additional nuclear suppression by about a factor two compared to the J/$\psi$ has been observed in Run 1,  cannot be explained by those models. One calculation tries to solve the puzzle by the introduction of late stage interactions of the resonant state~\cite{Ferreiro:2014bia}. It can describe the additional suppression within the current experimental precision. Another approach with late stage interaction is also able to describe the additional $\psi$(2S) suppression fairly well~\cite{Du:2015wha}. in Fig.~\ref{fig:pPb13}, the forward-backward ratio of $D^0$-meson production at $\sqrt{s_{NN}}=$ 5TeV is shown and compared with a calculation using the nuclear parton distribution function set EPS09. The approach describes well the data. This preliminary result~\cite{LHCb-CONF-2016-003} will be soon superseded by a publication exploiting the full data sample, corresponding to about a factor 10 more statistics compared to the analysis presented here.

\begin{figure}[t]
  \includegraphics[width=0.42\textwidth]{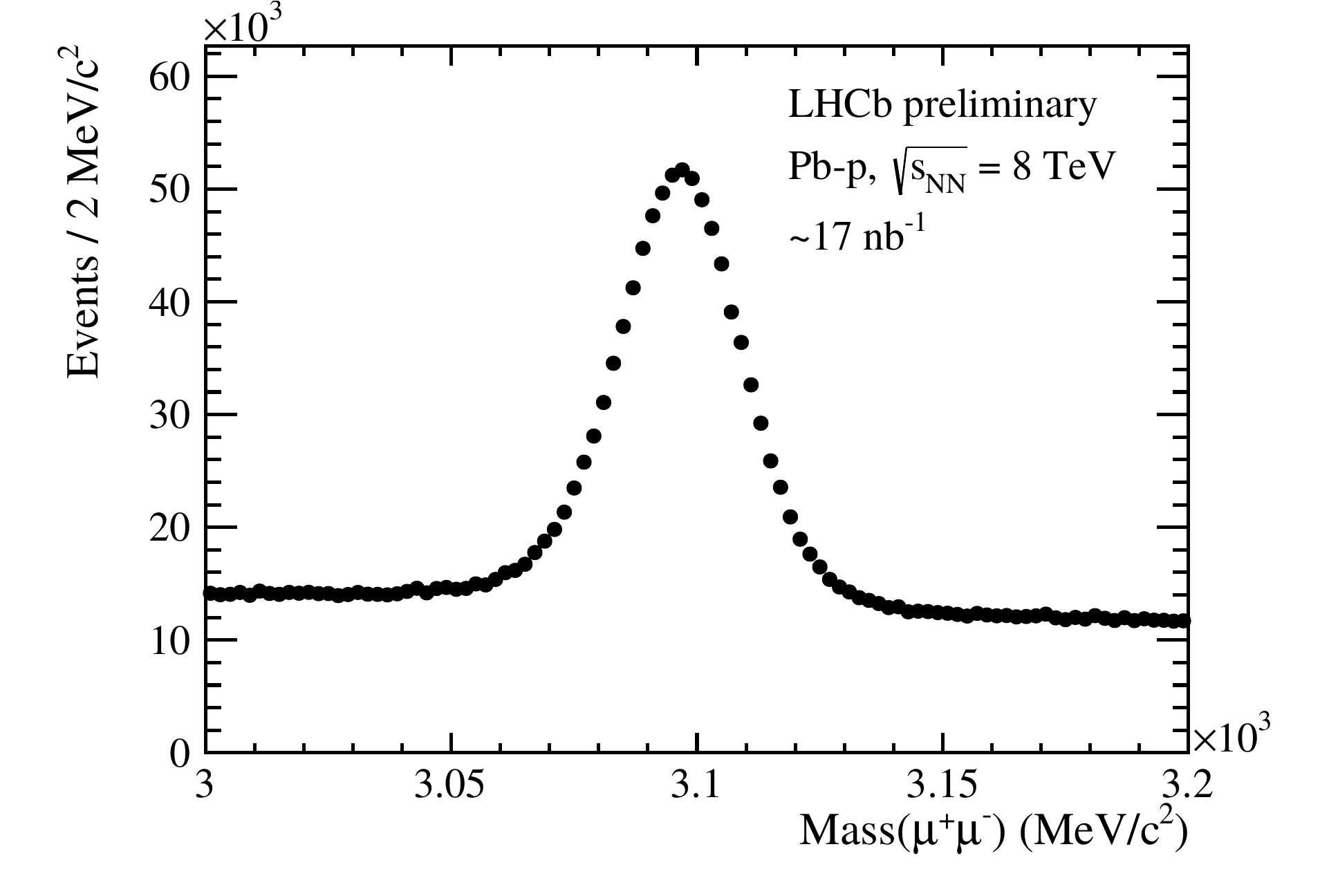}
  \includegraphics[width=0.42\textwidth]{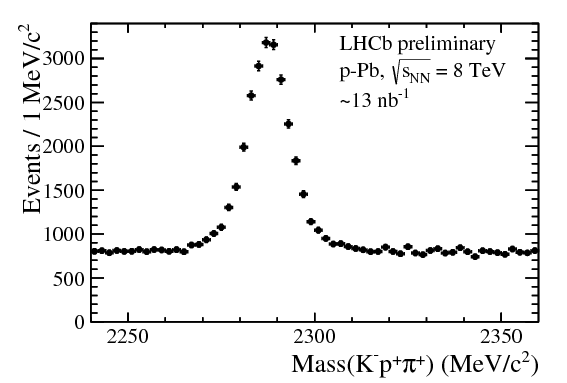}
  
  \includegraphics[width=0.42\textwidth]{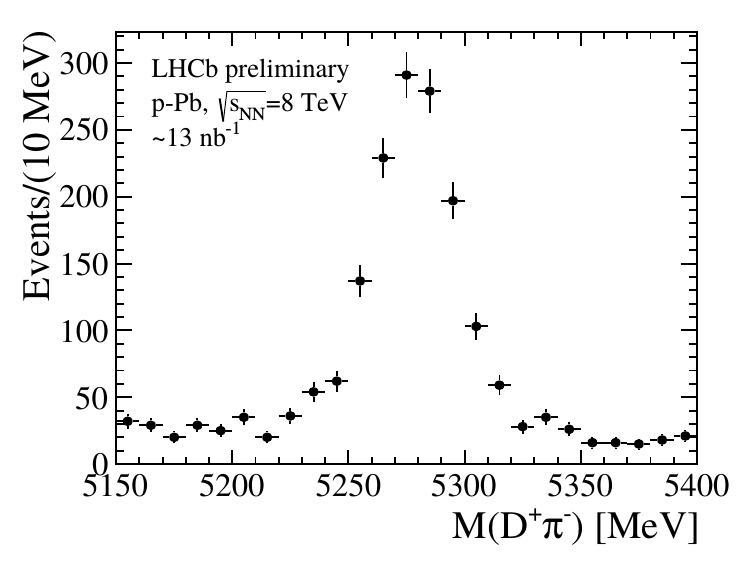}
  \includegraphics[width=0.42\textwidth]{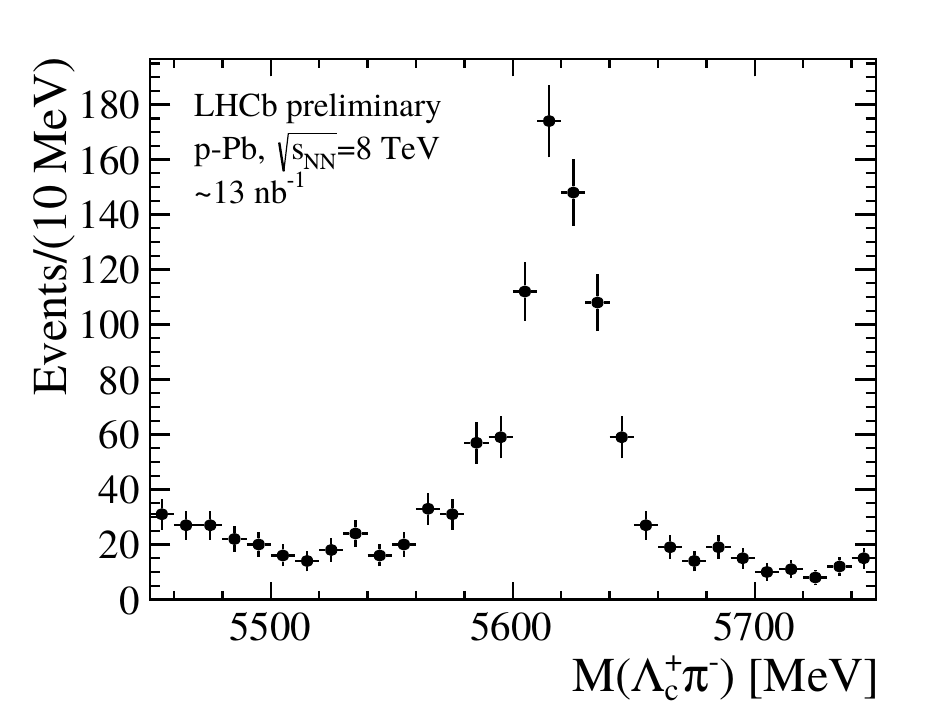}
  \caption{Invariant mass peaks from the 2016 $p$Pb data taking campaign: J/$\psi$ in Pb$p$ collisions (top left), $\Lambda_c$ in $p$Pb collisions, B$^0$ in $p$Pb (bottom left) and $\Lambda_b$ in $p$Pb (bottom right). }
  \label{fig:pPb16}
  \end{figure}

The vast majority of heavy-flavour observables with the 2013 $p$Pb data sample, the $\Upsilon$~\cite{LHCb-PAPER-2014-015} and $Z$-boson~\cite{LHCb-PAPER-2014-022} analyses even to a larger extent, are strongly statistically limited or not yet accessible. In 2016, the LHC provided an integrated luminosity larger by about a factor 10 for the $p$Pb (1.1 nb$^{-1}$ versus about 13 nb$^{-1}$) and a factor 30 for the Pb$p$ (0.5 nb$^{-1}$ versus about 17 nb$^{-1}$) beam configuration at $\sqrt{s_{NN}}=8.2$~TeV. These data samples will allow for more precise meausurements in the charm and beauty sector as the performance plots of various particle species depicted in Fig.\ref{fig:pPb16} already promise. In addition, it will be possible to add to the heavy-flavour programme significant measurements of electro-weak or electromagnetic probes as Drell-Yan production at low masses as well as  Z and W production. These theoretical clean measurements will allow to constrain further different possible nuclear modification mechanisms in the asymmetric $p$Pb collision system, see e.g. in Ref.~\cite{Arleo:2015qiv}.

The upgrade of the LHCb experiment is aiming at a continuous read-out of the detector at 40 MHz at around 5 times higher luminosities implying also 5 times larger number of simultaneous collisions per bunch crossing~\cite{LHCb-TDR-012}. The replacement of the strip VELO detector with a pixel detector and of all tracking stations with detectors with larger granularity~\cite{LHCb-TDR-013,LHCb-TDR-015} will be  beneficial for the heavy-ion programme of LHCb. 

LHCb allows to measure heavy-flavour production at forward rapidity providing unique coverage down to vanishing transverse momentum. Meausurements of open and closed heavy-flavour hadrons are possible in the same kinematic domain. The detector recorded large data samples in $p$Pb and Pb$p$ collisions with unprecedented opportunities and a non-negligible amount of minimum bias PbPb collisions  allowing unique studies in peripheral and semi-peripheral collisions.

\textit{Acknowledgement}: The contact author acknowledges support from the 
European Research Council (ERC) through the project EXPLORINGMATTER,
founded by the ERC through a ERC-Consolidator-Grant.






\vspace{-0.7cm}
\bibliographystyle{elsarticle-num}
\bibliography{LHCb-CONF,LHCb-DP,LHCb-PAPER,LHCb-Phys,LHCb-TDR}







\end{document}